\documentstyle[11pt,aaspp4]{article}

\slugcomment{}

\begin{document}

\title{Coevolution of Quantum Wave Functions and the Friedmann Universe\altaffilmark{1}}

\author{W. Q. Sumner}

\affil{Box 588, Kittitas, WA 98934 USA}

\author{D. Y. Sumner}

\affil{Department of Geology, University of California, Davis, CA 95616 USA}

\altaffiltext{1}{This is a copy of Sumner, W.Q., \& Sumner, D.Y.  2000, ``Nauka i Obrazovaniye,'' 4/5, 113-116. Gorno Altaysk, Russia}

\begin{abstract}

Erwin Schr\"odinger (1939) proved that quantum wave functions coevolve with the curved spacetime of the Friedmann universe.  Schr\"odinger's derivation explains the Hubble redshift of photons in an expanding universe, the energy changes of moving particles, and establishes the coevolution of atoms and other quantum systems with spacetime geometry.  The assumption often made that small quantum systems are isolated and that their properties remain constant as the Friedmann universe evolves is incompatible with relativistic quantum mechanics and with general relativity.

\end{abstract}

\section{Introduction}

Dirac (1958) summarized the prevailing paradigm of his time in relativistic quantum mechanics by writing,  ``There is no need to make the theory conform to general relativity, since general relativity is required only when one is dealing with gravitation, and gravitational forces are quite unimportant in atomic phenomena.''  In contrast, Schr\"odinger (1939) argued that if spacetime is curved as general relativity demands, then its effects on quantum processes must not be dismissed without careful investigation.  Using the equations of relativistic quantum mechanics, Schr\"odinger proved that quantum wave functions coevolve with the curved spacetime of the Friedmann universe.

Schr\"odinger found that the plane-wave eigenfunctions characteristic of flat spacetime are replaced in the curved spacetime of the Friedmann universe by wave functions that are not precisely flat and that have wavelengths that are directly proportional to the Friedmann radius.  This means that the eigenfunctions change wavelength as the radius of the universe changes and the quantum systems they describe follow.  In an expanding universe quantum systems expand.  In a contracting universe they contract.

From this quantum mechanical perspective Schr\"odinger confirmed the changes in both photon and particle momenta well known from general relativity, giving confidence in the logic of each approach and in the tie of Friedmann spacetime geometry to quantum processes.

These changes in quantum systems may equivalently be viewed as a logical consequence of the fact that the energy and momentum of an ``isolated'' system can change in general relativity when the spacetime geometry of the universe changes.  ``In an expanding space {\it all momenta decrease} \ldots for bodies acted on by no other forces than gravitation \ldots This simple law has an even simpler interpretation in wave mechanics: all wavelengths, being inversely proportional to the momenta, simply expand with space.'' (Schr\"odinger 1956).

Schr\"odinger's arguments are general and have applications beyond just photons and single particles.  His reasoning applies to all quantum systems.  Every quantum wave function coevolves with the Friedmann universe.

\section{General Relativity}

The equations of general relativity equate the curvature of spacetime to the properties of matter, \begin{equation} G_{\alpha\beta}={8}{\pi}T_{\alpha\beta}. \end{equation} The Einstein tensor $G_{\alpha\beta}$ is calculated from the metric tensor of spacetime geometry, $g_{\alpha\beta}$, and its derivatives.  $T_{\alpha\beta}$ is the energy-momentum tensor of the system under consideration.

Assuming a spatially isotropic and homogeneous energy distribution for the universe, Friedmann (1922) found two solutions to these equations.  The open solution is infinite and expands forever.  The other, the closed solution, is finite and expands to a maximum size and then contracts.  Schr\"odinger chose Friedmann's closed solution to model the universe, an assumption whose validity will be confirmed below.  Using three angles $\chi,\vartheta,\varphi$ as spatial coordinates, the Friedmann line element can be written (Schr\"odinger 1939)\begin{equation}ds^2\,=\,c^2 dt^2\,-\,R^2 (t)\left[d\chi^2\,+ \,sin^2\chi\left(d\vartheta^2\,+\,sin^2\vartheta \, d\varphi^2\right)\right]. \end{equation} $R(t)$ is the radius of spatial curvature, and $t$ is time.

Using this spacetime geometry, general relativity proves that photon wavelengths are proportional to  $R(t)$.  In an expanding universe, photon wavelengths increase with time and in a contracting universe, photon wavelengths decrease.  The evolving spacetime geometry of the universe changes the energy and momentum of individual photons.  While the resulting redshift of a photon in an expanding universe is often referred to as a D\"oppler shift, it really isn't in the special relativistic sense since the shift depends only on the radii $R(t)$ of the Friedmann geometry at the times of emission and of observation and does not depend on $dR/dt$ at either of those times.

A common interpretation of general relativity is that matter determines spacetime curvature. For example, one thinks of the mass of the earth as determining the local curvature of spacetime geometry (i.e. gravity).  But the opposite view is equally valid: Spacetime geometry also defines matter.  The evolution of Friedmann geometry redefining photon energy is an example.

The decrease of energy and momentum of free particles in an expanding Friedmann universe and their increase in a contracting one is also well known from general relativity.  The derivation for free particles, like the one for photons, uses only the characteristics of the geodesics of Friedmann geometry (Tolman 1934).  The decrease in energy in an expanding universe is not limited to photons and particles.  It has also been shown that electromagnetic fields lose energy in an expanding universe (M{\o}ller 1952, Sumner 1994), a mathematical result consistent with the loss of electromagnetic field energy in redshifted photons.

There are no universally valid energy and momentum conservation laws in general relativity.  The energy that was attributed to gravity in Newtonian mechanics is missing, reflected instead by changes in spacetime curvature.

\begin{quote}\ldots we may assert that the whole universe loses energy \ldots where does this energy go to? In Newtonian mechanics one would say that it is spent to overcome the mutual gravitational attraction and stored as potential energy of gravity.  From Einstein's theory the notions of gravitational pull and potential energy have disappeared, though they are used occasionally for brevity of speech.  But are there not very important conservation laws at the basis of this theory, including the conservation of energy?  Are they not violated if energy is said to diminish without there being a flow outwards (which there cannot be, because there is no boundary and no outside)?  Well, no.  The conservation law does not allow one to assert that the energy content of any given spatial region is constant provided there is no energy traffic through the boundary; and that for the simple reason that the energy density is a tensor component (not an invariant) and its integral over an invariantly fixed region of space has no invariant or covariant meaning at all, not even that of a tensor--or vector--{\it component } (Schr\"odinger 1956).\end{quote}

The assessment that changing spacetime geometry in general relativity alters energy and momentum is mathematically convincing.  Since energy and momentum changes imply that quantum wave functions change, it should come as no surprise that relativistic quantum mechanics gives precisely the same results as those given by general relativity.

\section{Quantum Mechanics}

Every mathematical model of every physical system is simpler than reality.  While simplifications are essential to make mathematics manageable, there are always the questions of what can be acceptably simplified.

Consider a small quantum system in a Friedmann universe.  Can one assume that the system is isolated from the rest of the universe?  Maybe or maybe not.  The fact that spacetime geometry at every point depends on the rest of the universe means that no quantum system is truly isolated.  Does that mean it is always unreasonable to assume that small, isolated quantum systems exist and study them mathematically?  Certainly not.  What it does mean is that one must be ready to carefully explore the implications of the simplifying assumption that quantum systems are isolated as Schr\"odinger did.

For a relativistic wave equation for flat spacetime, Schr\"odinger (1939) used  \begin{equation} \nabla^2\psi-\frac{1}{c^2}\frac{\partial^2\psi}{\partial t^2}-\mu^2\psi=0, \end{equation} where $\psi$ is the wave function, $\mu\,=\,mc/\hbar$, and $m$ is the rest mass of the particle.  For a general spacetime metric, $g_{\alpha\beta}$, this equation becomes \begin{equation} \frac{1}{\sqrt{-g}}\frac{\partial}{\partial x^\alpha}\left[ \sqrt{-g}\, g^{\alpha\beta}\frac{\partial\psi}{\partial x^\beta} \right]+ \mu^2\psi=0, \end{equation} where $g$ is the determinant of $g_{\alpha\beta}$ and $x^\alpha$ are generalized coordinates.

To describe the evolving geometry of the universe, Schr\"odinger chose Friedmann's closed solution to the equations of general relativity, equation (2).  Substituting this into equation (4) gives \begin{equation} -R^{-2}\,K[\psi] \,+\,\frac{1}{c^2} R^{-3}\frac{\partial}{\partial t}\left[R^3 \frac{\partial\psi}{\partial t}\right]\,+\,\mu^2\psi\,=\,0. \end{equation} $K[...]$ is a second order differential operator whose eigenfunctions are spherical harmonics with the eigenvalues $-n(n\,+\,2)$, $n\,=\,0,\,1,\,2,\,3,\ldots$  Equation (5) can be solved by the method of separation of variables. By letting \begin{equation} \psi(\chi,\vartheta,\varphi,t)\,=\,\omega(\chi,\vartheta,\varphi)\, f(t), \end{equation} Schr\"odinger showed that $\omega$ is an eigenfunction of $K$. For the extremely large eigenvalues necessary to obtain wavelengths of atomic dimensions, Schr\"odinger (1939, 1956) proved that $\lambda(t)$, the wavelength of $\omega$, is proportional to the radius of curvature of the universe, \begin{equation} \lambda(t)\,\alpha\,R(t). \end{equation} The dependence of wavelength on spacetime curvature may be viewed as a direct result of there being an integral number of nodes of $\omega$  on the circumference $2\pi R(t)$ of the Friedmann space.  As $R(t)$ changes and the number of nodes stays the same, the wavelength of $\omega$ must change as described by equation (7).  It is clear from this derivation that changes in quantum eigenfunctions are tied to the changes in spacetime curvature measured by $R(t)$ and have nothing to do with D\"oppler shifts of moving objects.

The complete set of solutions that satisfy equation (5) can be used to describe arbitrary quantum wave functions in Friedmann geometry.  The evolution of eigenfunctions requires that the quantum wave functions described by them evolve in exactly the same way.

\section{Photons and Particles}

A photon, with a wavelength $\lambda(t)$, is composed of waves very close to one another in wavelength and propagation direction.  Schr\"odinger (1939) showed that $\lambda(t)$ is proportional to the Friedmann radius $R(t)$, \begin{equation} \frac{\lambda (t_2)}{\lambda (t_1)}=\frac{R(t_2)}{R(t_1)}. \end{equation} The wavelength of a photon evolves because its constituent eigenfunctions evolve with the Friedmann universe. Equation (8) is identical to the equation derived in general relativity to explain Hubble redshift (Hubble \& Tolman 1935).

The evolution of eigenfunction wavelengths applies equally to particles since equation (7) holds true for the eigenfunctions of both light ($\mu=0$) and matter ($\mu\not=0$).   Schr\"odinger (1939) showed that a particle's momentum is proportional to $R^{-1}(t)$ and that its de Broglie wavelength is proportional to $R(t)$.  In an expanding universe, a particle slows down and becomes larger. This change in particle momentum with spacetime curvature derived by Schr\"odinger reproduces another well-known result of general relativity (Tolman 1934). Schr\"odinger (1939) also showed that the volume $V$ of a particle, calculated from the spread in its wave packet, depends on the radius of curvature, \begin{equation} V\,\alpha\,R^3(t).\end{equation}

In deriving these results, Schr\"odinger used two simplifications.  He assumed that the wavelengths of the eigenfunctions are much smaller than $R(t)$ and that they change much faster than $R(t)$, making a classic argument for adiabatic change. These assumptions are extraordinarily good except in the very high curvature regions near spacetime singularities. For mathematical convenience, Schr\"odinger derived the change in wavelength for spin zero particles. Identical results are obtained for other spin states (Schr\"odinger 1939), because each component of the wave function for a particle with non-zero spin satisfies an equation with similar form (Berestetskii 1982).

One important distinction between photons and particles is the different relationship between momentum and energy they have.  For photons, wavelength is inversely proportional to momentum which is proportional to energy.  For particles, wavelength is also inversely proportional to momentum but energy is proportional to the {\it square} of the momentum.  This means that as the universe expands or contracts the energies of photons and the energies of particles change at different rates.

\section{Atoms}

Atoms are described by quantum wave functions and so coevolve with the Friedmann universe. $\bar{r}(t)$, the mean radius of an atom calculated from its wave function, is proportional to the radius of spacetime curvature, \begin{equation} \bar{r}(t)\,\alpha\,R(t). \end{equation}

This change is identical to the change in photon and particle wavelengths and occurs for precisely the same reasons.  The change in the mean radius is consistent with Schr\"odinger's results for the change in the volume of a particle, equation (9), and for the change in a particle's de Broglie wavelength.

This view of atomic change comes from considering the atom as a ``particle'' and does not depend on the details of the atomÕs internal structure.  Another perspective on atomic change comes from peering ``inside'' and looking at the electrical field that binds the electron to the nucleus.

As was noted above, electromagnetic energy changes with spacetime geometry.  These changes in energy are reflected by changes in the strength of the electrical field.  Changes in the electrical fields that bind electrons in atoms imply changes in mean atomic radii.  The dependence of atomic radii on Friedmann geometry found in this way is exactly the same as the one established here (Sumner 1994).

Evolution of atomic wave functions also implies that electron energy levels, $E_n$, change with $R(t)$.  For the hydrogen atom, the dependence of $E_n$ on $R(t)$ is \begin{equation} E_n\,\alpha\,\frac{1}{R^2(t)n^2},\end{equation} where $n$ is the principle quantum number (Sumner 1994).  Since electron energy levels determine the wavelengths of atomic emissions, atomic spectra evolve as well. Thus, the hydrogen alpha emission line changes color as the size of the universe changes, becoming redder in an expanding universe.

The rate of atomic evolution depends on the rate of change of spacetime curvature, which is measured by the Hubble constant.  A Hubble constant of $90 \,kms^{-1}Mpc^{-1}$ implies a Hubble time of about $10^{10} years$, so $R(t)$ and atomic properties change roughly one part in $10^{10}$ per year. The relative importance of atomic changes of this magnitude depends on the system being considered. For many quantum problems, the effects of changes in spacetime curvature of this size are unimportant: Dirac's {\it a priori } hypothesis that ``gravitational forces are quite unimportant in atomic phenomena'' is appropriate and it is reasonable to assume isolation.

However, for matters of principle and for quantum processes that span long periods of time, the coevolution of atomic wave functions and the Friedmann universe is important.  Meter sticks change size as constituent atoms evolve, fundamentally changing the way experimental length is correlated to mathematical distance.  In the case of atomic emissions long ago from distant galaxies, the evolutionary change in atoms during the time it has taken for the photons to reach the earth is enough to reverse the interpretation of Hubble redshift to imply that the universe is closed and is presently contracting (Sumner 1994).  This result justifies Schr\"odinger's original decision to used closed Friedmann geometry to describe the universe.

The reversal in interpretation of redshift is a direct result of the different rates of change of energy for photons and atoms.  In an expanding universe photons are simply out redshifted by atoms.  A relative blueshift would be observed if a redshifted photon was compared against an atom that had redshifted even more.  Only when photons and atoms are both blueshifted in a contracting universe is the observed shift red.  Hubble redshift proves that the universe is contracting and closed.

\section{Other Quantum Systems}

There are two complimentary ways of studying the evolution of quantum systems.  One is to view the quantum system as a particle and study the evolution of its wave function.   The other is to seek insight into how its internal interactions evolve.  The overall view is straight forward, while understanding evolutionary changes in specific interactions can be more difficult.  The two views of atomic change, as atomic wave functions evolving with Friedmann geometry or as a result of the evolution of electrical fields, illustrate these two different approaches.

The concept of particle is used in quantum mechanics for something whose internal structure is unimportant to the problem being considered.  Thus, a ``particle'' may be an electron, a nucleon, an atom, a molecule, or something much larger, depending on the circumstances.   If internal structure cannot be ignored, the simplifying assumption of the particle view must be abandoned.

If a quantum system consists of many particles, it is usually assumed that a complete set of dynamical variables which describes the behavior of each particle can still be used, even when there are interactions between the particles.  This implies that the wave function for the complete system can be written as a sum of the products of the various possible wave functions of the individual particles.  If there are interactions, a relativistic quantum mechanical description of the interactions must be available to fully understand the effects of the evolution of Friedmann spacetime on the quantum system.  If individual particles within the system do not interact the problem is trivialized--one just considers the individual particles.

An assumption critical to this paper is that spacetime geometry can be approximated by the Friedmann metric, equation (2).  It is an extremely accurate approximation for most problems.  However, for some astrophysical situations where quantum systems have very high masses and very high mass densities, it is inappropriate.   For those cases, a more adequate metric must be determined and the relationship between quantum wave functions and spacetime curvature explored anew.

\section{Conclusions}

Schr\"odinger's proof of the evolution of quantum wave functions is straight forward and compelling. If one accepts the fundamental tenants of relativistic quantum mechanics and general relativity, one has no choice but to accept the coevolution of quantum wave functions and the Friedmann universe.

\end{document}